\documentclass[%
 reprint,
superscriptaddress,
showpacs,
showkeys,
 amsmath,amssymb,
 aps,
 pre,
]{revtex4-1}

\usepackage{graphicx}
\usepackage{dcolumn}
\usepackage{bm}
\usepackage{siunitx}		
\renewcommand{\vec}[1]{\bm{#1}} 		
\newcommand{\uvec}[1]{\hat{\bm{#1}}} 	

\usepackage{picture}			
\usepackage{commath}		
\usepackage{todonotes}

\begin{document}

\graphicspath{{./pix/}} 

\preprint{APS/123-QED}

\title{Molecular simulations of entangled defect structures around nanoparticles in nematic liquid crystals}
\author{Anja Humpert}
\email{anja.humpert@gmail.com}
\author{Samuel F.\ Brown}
\email{s.f.brown@warwick.ac.uk}
\affiliation{Department of Physics, University of Warwick, Coventry CV4 7AL, United Kingdom}
\author{Michael P.\ Allen}
\email{m.p.allen@warwick.ac.uk,m.p.allen@bristol.ac.uk}
\affiliation{Department of Physics, University of Warwick, Coventry CV4 7AL, United Kingdom}
\affiliation{H. H. Wills Physics Laboratory, Royal Fort, Tyndall Avenue, Bristol BS8 1TL, United Kingdom}

\date{\today}

\begin{abstract}
We investigate the defect structures forming around two nanoparticles 
in a Gay-Berne nematic liquid crystal
using molecular simulations.
For small separations, disclinations entangle both particles 
forming the figure of eight, the figure of omega and the figure of theta.
These defect structures are similar in shape and occur with a comparable frequency to micron-sized particles
studied in experiments.
The simulations reveal fast transitions from one defect structure to another 
suggesting that particles of nanometre size cannot be bound together effectively.
We identify the `three-ring' structure observed in previous molecular simulations
as a superposition of the different entangled and non-entangled states over time 
and conclude that it is not itself a stable defect structure.
\end{abstract}


\maketitle



\section{\label{s.intro}Introduction}

\begin{figure}
\centerline{\includegraphics[width=0.95\columnwidth]{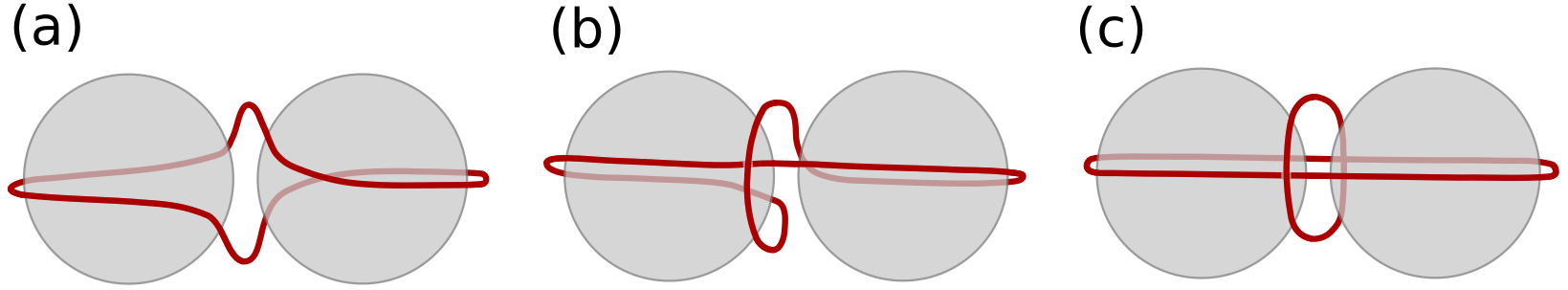}}
\caption
{Schematic of two nanoparticles (grey) in nematic host (not shown) entangled by a disclination line (dark red) forming the 
(a) figure of eight $\infty$, (b) the figure of omega $\Omega$ and (c) the figure of theta $\Theta$.
\label{fig:sketch_entangled}}
\end{figure}

Spherical particle inclusions in liquid crystals distort the orientational order of the surrounding molecules
inducing topological defects.
Defect regions are defined by the absence of orientational order, i.e.\ the liquid crystal is melted.
Such defects arise due to the competition between liquid crystal molecules trying to align along
the average molecular direction, called the director, and trying to fulfil the surface anchoring condition of the nanoparticle.
For small particles or particles confined to a thin nematic cell,
the defect region forms a defect ring near the equator of the particle with respect to the director, 
commonly referred to as a Saturn ring \cite{Stark2002}.
Particles surrounded by a Saturn ring show quadrupolar interactions
due to the symmetry of the director field surrounding them;
hence such particles are called quadrupoles.
For two quadrupoles in close vicinity, entangled defects were found to spontaneously arise 
when the surrounding nematic is distorted.
Using laser tweezers, which allow easy manipulation with great precision, 
a range of reproducible entangled objects have been found \cite{Guzman2003,Araki2006,Ravnik2007,Skarabot2008,Ravnik2009,Copar2011,Kim2004}.
Here `entangled' means that both particles are surrounded by a single defect line, 
commonly referred to as a disclination line.
The three main structures comprise the figure of eight ($\infty$), the figure of omega ($\Omega$) and the figure of theta ($\Theta$).
Schematics are shown in Fig.~\ref{fig:sketch_entangled}.
Micrographs of these entangled defect structures observed in experiments can be found in Fig.~4 of \citet{Tkalec2013}.
For the figure of eight defect a single disclination line winds around the particles 
in the manner shown in Fig.~\ref{fig:sketch_entangled}(a).
The figure of omega consists of one disclination line that surrounds both particles near the equator,
entering into the space between the particles 
to descibe a shape similar to the Greek letter $\Omega$.
The figure of theta consists of a disclination line surrounding both particles 
and an additional defect loop in the plane in between the particles. 
Note the spontaneous symmetry breaking for the first two of these structures due to their chirality.
The two states are degenerate and hence left-handed and right-handed structures occur with the same frequency.

These entangled defects were observed in experiments and predicted numerically for micron-sized particles
using phenomenological mean-field calculations based on Landau-de Gennes free energy minimisation \cite{Ravnik2009a,Ravnik2009}.
At first glance these entangled structures appear to disobey the restriction that the net topological charge within the liquid crystal must be zero;
however \citet{Copar2011} resolved this apparent contradiction with the introduction of a self-linking number.

A closer look at the different defect structures reveals that entangled defects differ only within a tetrahedral region.
Theoretically, by rotating the tetrahedron, one entangled state can be transformed into another \cite{Copar2011}.
In practice, this is equivalent to cutting and reconnecting the disclination lines,
whereas in experiments this rewiring is achieved by locally applying laser tweezers
to heat the region around the tetrahedron.

Molecular simulations of two quadrupoles in close vicinity
have shown a three-ring structure for particle inclusions of nanometre size \cite{Grollau2003,Kim2004,Guzman2003}.
This structure is similar to the figure of $\Theta$;
however the inner loop connects with the outer loop at two nodes.
Two possible explanations have been proposed as to why the structure
differs from the ones observed in experiments.
Earlier studies declared it a transient state \cite{Araki2006}.
It was later proposed \cite{Hung2009} 
that instead the size of the particles is crucial and that the three-ring structure appears
for particles of nanometre scale, 
whereas micron-sized particles form only the three different entangled defects ($\infty$, $\Theta$ and $\Omega$).
In this paper we propose an alternative explanation for the three-ring structure.

The system of two entangled particles can be thought of as a building block for more complex systems.
Recent work has shown one-dimensional \cite{Ravnik2007} as well as two-dimensional \cite{Ravnik2009d,Araki2013,Tkalec2011} aggregates, where the disclination line winds around all the particles.
Entanglement has also been observed for microspheres and microfibres \cite{Nikkhou2015a}.
When bringing many particles into close vicinity, knotted structures can be created \cite{Jampani2011,Tkalec2011,Copar2015}.
By applying the laser tweezers in unknotted regions, additional knots can be created and vice versa.
\citet{Wood2011} showed that liquid crystals with a high number of nanoparticle inclusions can be used 
to synthesise a soft solid, which shows high rigidity due to a network of entangled defect lines.
These materials have important features for potential use in biosensors \cite{Agarwal2008}.

Entangled and knotted systems can be manipulated not only by the use of laser tweezers,
but also by using strong local electric fields, hydrodynamic flow, temperature changes
and the use of different boundary conditions and confined geometries.
This allows the creation and modification of defects in liquid crystals in a rather controllable way
and hence a variety of complex systems can be designed.

In this paper we outline a model to efficiently simulate entangled nanoparticles.
Simulation details are given in Section \ref{s.model}.
In Section \ref{s.sausage} we present how the different defect types are automatically categorised.
The results, in particular the time scales and dynamics of transitions from one entangled defect structure to another,
are discussed in Section \ref{s.rd}. 
In addition, the origin of the three-ring structure, the only structure observed in previous molecular simulations, is identified.
Conclusions are drawn in Section \ref{s.conclusions}


\section{Model and simulation details}\label{s.model}
The entanglement was studied by simulating several systems with nanoparticle inclusions
of various sizes and separations.
The nematic host was simulated using the well-known soft Gay-Berne potential 
(see appendix \ref{appendix.gb}),
a coarse-grained single-site potential that represents the interaction energies between two elongated particles.
We chose a length-to-width ratio of $\kappa=3$. 
The remaining Gay-Berne coefficients were set to $\mu=1$, $\nu=3$ and $\kappa'=5$ 
and a potential cutoff of $5\sigma_0$ was applied.
Note that reduced units were used throughout by setting $\sigma_0=\epsilon_0=1$
and the molecular mass $m_0=1$ leading to a basic unit of time $\tau_0=\sigma_0\sqrt{m_0/\epsilon_0}$.
The moment of inertia was set to $I=0.5m_0\sigma_0^2$ assuming uniform mass distribution.
The initial director orientation was chosen to lie along the $z$ direction.

Spherical particle inclusions of radius $R_\text{NP}$ were placed in the box,
interacting with the liquid crystal molecules via a purely repulsive Lennard-Jones potential (see appendix \ref{appendix.homeo}).
Their positions were fixed throughout the simulation.
The system was simulated for particles separated along $x$ with a surface-to-surface separation of $\Delta$.
Five different systems of nematics with nanoparticle inclusions 
of radii $R_\text{NP}=10\sigma_0$, $15\sigma_0$ and $20\sigma_0$ were studied.
In all cases,
box dimensions were chosen to give a bulk density $\rho\sim0.3$,
and the temperature was chosen to be $T=3.0$,
which, for this system,
gives a nematic liquid crystal phase.
Simulation details are given in the following subsections.

\subsection*{Nanoparticles of radius $R_\text{NP}=10\sigma_0$}
Two spherical particle inclusions of radius $R_\text{NP}=10\sigma_0$ were placed in the box,
separated along $x$, with $\Delta = 2.94\sigma_0$ as well as $\Delta = 10.3\sigma_0$.
A cubic simulation box of length $\sim120\sigma_0$ with periodic boundaries was used.
The large simulation box is necessary to avoid any interference between nanoparticles through the boundaries.
The system was equilibrated over \num{1.6e6} time steps using an $NpT$ ensemble,
followed by a production run,
also in the $NpT$ ensemble, of \num{2.5e5} time steps
with a time step $\Delta t=0.004$.
Molecular positions and orientations were stored every \num{500} time steps.
The total number of Gay-Berne molecules was \num{512000}.

\subsection*{Nanoparticles of radius $R_\text{NP}=15\sigma_0$}
The simulation was repeated with particles of size $R_\text{NP}=15\sigma_0$
separated along $x$ with $\Delta = 5\sigma_0$ as well as $\Delta = 10\sigma_0$.
Most other simulation details were unchanged.
Two modifications were made to reduce computational cost.

\begin{figure}
\includegraphics[width=0.99\columnwidth]{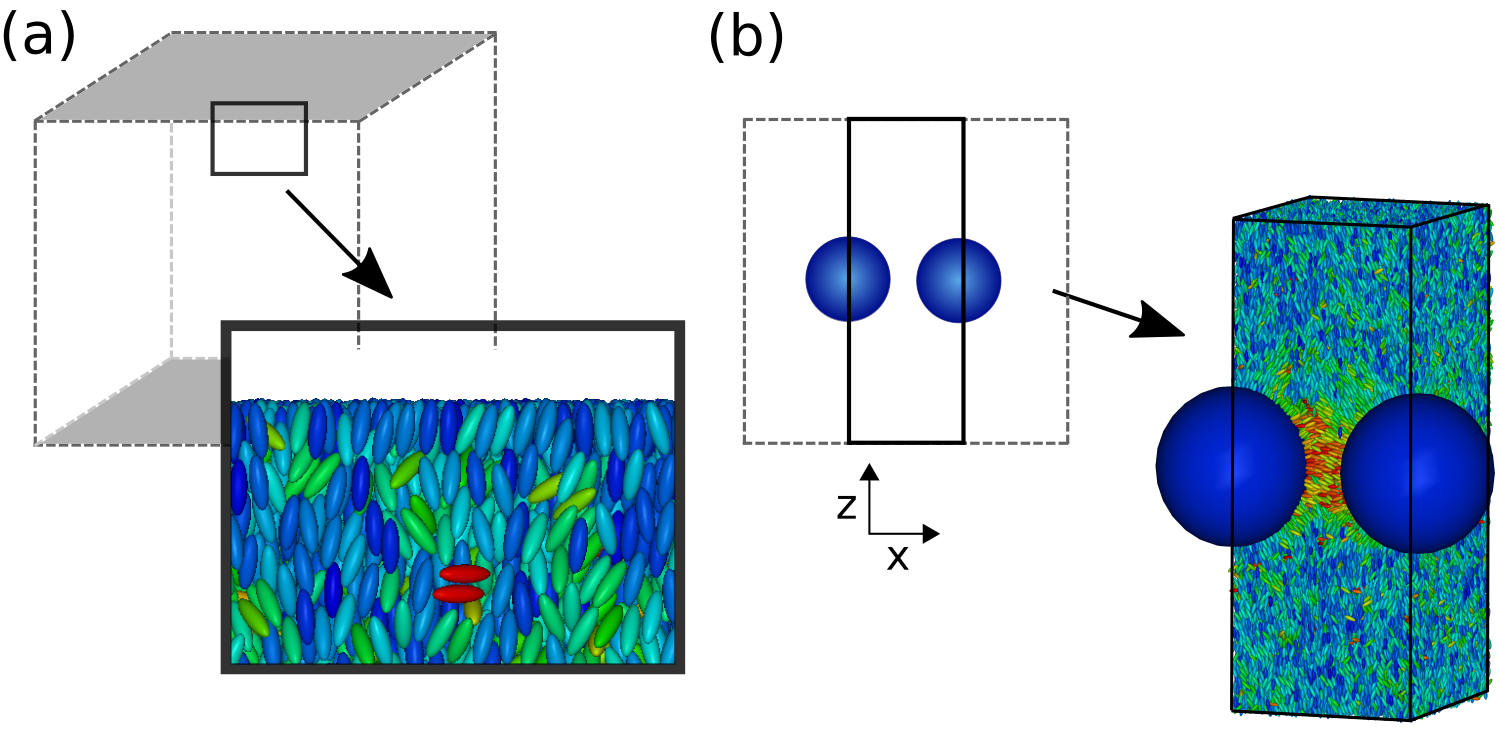}
\caption{Schematic of simulation box. 
(a) Typical snapshot of Gay-Berne molecules near a Lennard-Jones wall
applied near the $z$ boundaries indicated in grey.
Gay-Berne molecules were colour-coded according to their orientation with respect to the director.
(b) Two-dimensional view of the simulation box (grey dashed region) in the $x$-$z$ plane.
The black box indicates the area used to simulate an infinitely long one-dimensional chain of nanoparticles (blue).
The box length along $y$ was unchanged.
Molecular snapshot shows a slice of the new simulation box with the nanoparticles at its centre.
\label{fig:simulation_setup}}
\end{figure}

The first modification consists of the introduction of flat walls at the $z$ boundaries.
A simple Lennard-Jones 12-6 function of the $z$ coordinate, relative to the wall, was utilised with ${\epsilon_0}={\sigma_0}={1}$ and a potential cutoff of $2.5\sigma_0$.
This has the advantage that $L_z$ can be chosen to be smaller
than in the system described above, without the danger of long-range interactions 
between images of nanoparticles across the boundaries;
this type of boundary also stabilizes the bulk director in the $z$ direction.
$L_z$ must be sufficiently large to accommodate the density and order parameter
fluctuations near the walls, ensuring bulk behaviour in the centre of the simulation box.
A system snapshot of Gay-Berne molecules near a Lennard-Jones wall can be seen in Fig.~ \ref{fig:simulation_setup} (a).
One can clearly see layering near the wall that disappears quickly further away from the wall.

The second modification consists of the reduction of 
the simulation box length $L_x$ to $2R_\text{NP} + \Delta$
(so $L_x= 35 \sigma_0$ for $\Delta=5 \sigma_0$
and $L_x=40\sigma_0$ for $\Delta=10 \sigma_0$)
and the placement of one nanoparticle at ($\pm \frac{1}{2}L_x$,0,0),
illustrated in Fig.~\ref{fig:simulation_setup}(b).
With periodic boundary conditions along $x$
this represents an infinite chain of nanoparticles along the $x$ direction 
($L_y$ remains large enough to avoid periodic self-interaction).
This allows the reduction of the total number of Gay-Berne molecules by over 50$\%$;
choosing $L_y=L_z\sim 136\sigma_0$ in both cases,
with $N=\num{182000}$ for $\Delta=5 \sigma_0$ and $N=\num{210000}$ for $\Delta=10 \sigma_0$,
gave a bulk density $\rho\sim 0.3$.
The system was equilibrated over \num{7.5e5} time steps ($\Delta t=0.004$) using an $NVT$ ensemble,
followed by a production run ($NVT$ ensemble) of \num{4e5} time steps.
Molecular positions and orientations were stored every \num{500} time steps.

\subsection*{Nanoparticles of radius $R_\text{NP}=20\sigma_0$}
In this simulation the nanoparticle radius was increased to $R_\text{NP}=20\sigma_0$.
Other system parameters were chosen as for the system with $R_\text{NP}=15\sigma_0$.
No wall potential was applied and instead periodic boundaries were used across $z$.
The total number of Gay-Berne molecules was $N=\num{214000}$ with a moment of inertia of $I=2.5m_0\sigma_0^2$.
The box dimensions were chosen to be $L_x=50\sigma_0$ and $L_y=L_z \sim 120 \sigma_0$;
hence the separation between neighbouring nanoparticles was $\Delta = 10\sigma_0$.
The system was equilibrated over \num{8e5} time steps ($\Delta t=0.004$) using an $NVT$ ensemble,
followed by a production run ($NVE$ ensemble) of \num{6.5e5} time steps.
Molecular positions and orientations were stored every \num{500} time steps.


\section{Automated characterisation of defect types}\label{s.sausage}
To identify defect regions for the system snapshots stored,
a weighted order tensor was calculated following the approach suggested by \citet{Callan-Jones2006}.
This quasi-continuous tensor is given by
\begin{equation}
D_{mm'}(\vec{r}) = \frac{1}{N(V_{s})}\sum_{i\in V_{s}} w(|\vec{r}_i-\vec{r}|)u_{m}^{i}u_{m'}^{i}  \:,
\end{equation}
where $N(V_{s})$ is the number of molecules in the spherical sampling volume $V_{s}$
with radius $r_k$ centred at $\vec{r}$,
$\vec{r}_i$ is the position vector of particle $i$ and $u^{i}$ is the component of its orientation vector with $m,m' = x,y,z$.
$w(|\vec{r_{i}}-\vec{r}|)$ is a weighting function with the constraint
\begin{equation}\label{eqn:w}
\sum_{i\in V_{s}} w(|\vec{r_{i}}-\vec{r}|) = 1 \:.
\end{equation}
We choose a cubic b-spline,
which is a piecewise continuous cubic polynomial approximation to a Gaussian function \cite{Bartels1987},
scaled to obey Equation~\eqref{eqn:w}:
\begin{equation}
w(r_w) = \left\{
  \begin{array}{l l}
    \frac{1}{6}(3|r_w|^{3}-6r_w^2+4) & \quad 0\leq|r_w|\leq1 \\
    \frac{1}{6}(2-|r_w|)^{3} & \quad 1\leq|r_w|\leq2 \\
    0 & \quad \text{else,}  
  \end{array} \right.
\end{equation}   
where $r_w=2|\vec{r_{i}}-\vec{r}|/r_{k}$. 
$w$ is zero if $|\vec{r}_i-\vec{r}|$ is greater than the radius $r_{k}$.
$r_{k}$ was chosen to be 7.3$\sigma_0$, 
which corresponds to having roughly 30 Gay-Berne molecules inside the sampling volume $V_{s}$.
The eigenvalues of $D$ are labelled $\lambda_1 \ge \lambda_2 \ge \lambda_3$.
The uniaxial order of a grid point is defined as its alignment to the local nematic director field, and is given by the Westin metric 
$c_{l} =\lambda_{1}-\lambda_{2}$,
which is zero in the isotropic region and unity in perfectly ordered regions.
Within the bulk $c_l$ undergoes small variations, 
whereas significant reductions indicate defect regions.
We identified $c_l<0.05$ as an appropriate threshold to determine defect core regions.

For each system snapshot the weighted order tensor $D$ was calculated
on a regular 3D grid with a spacing of $0.25\sigma_0$
and defect regions were evaluated, which allowed the subsequent determination of the defect structure.
Due to the high number of snapshots this process was automated 
and the defects classified as follows.

All grid points of the defect core, corresponding to $c_{l} < 0.05$, were identified and stored;
all points with values above the threshold were discarded
We did not further distinguish between different values of $c_{l}$.

To separate points into disclination lines, the flood-fill algorithm was used.
To begin, an arbitrary starting point is chosen and added to a queue.
The algorithm then consists of removing the first point from the queue, adding all its neighbours to the queue, then adding the removed point to a list $L_i$.
This step is repeated until the queue is empty, at which point the list $L_i$ corresponds to a discrete contiguous volume of points.
A new starting point is then chosen to form the seed of a new list $L_j$, and the process continues until all points have been assigned to a disclination line.

Note that since only defect regions were stored, a point may not have neighbours in all six directions.
To efficiently implement the flood-fill algorithm for such an irregular array, a linked-list connectivity graph was constructed such that each point stores references to its neighbours.
Since disclination lines can only exist in the form of closed loops (or lines across periodic boundaries),
each defect line was checked for this criterion.
For coarser grids, noise artefacts in the data sometimes caused hairline breaks in the disclination lines, causing them to have `endpoints'.
These endpoints could be identified by examining the neighbour-to-neighbour distance between points, estimated using the all-pairs shortest-path Floyd-Warshall algorithm \cite{Cormen2001} to find the two points separated by the longest path.
Two endpoints in close proximity, and with no other endpoint nearby, could be automatically joined and their lists merged.

While the Floyd-Warshall algorithm gives a deterministic measure of distance along the disclination line,
it scales with the cube of the number of points $P$ and would be too slow for larger systems.
The sparsity of the point-connectivity graph would make a per-point Dijkstra approach more promising in these cases, with $\mathcal{O}(P^2 \log P)$ scaling \cite{Cormen2001}.
Coarse-graining could also be used to reduce the effective $P$.

\begin{figure}
\begin{center}
\includegraphics[width=0.5\columnwidth]{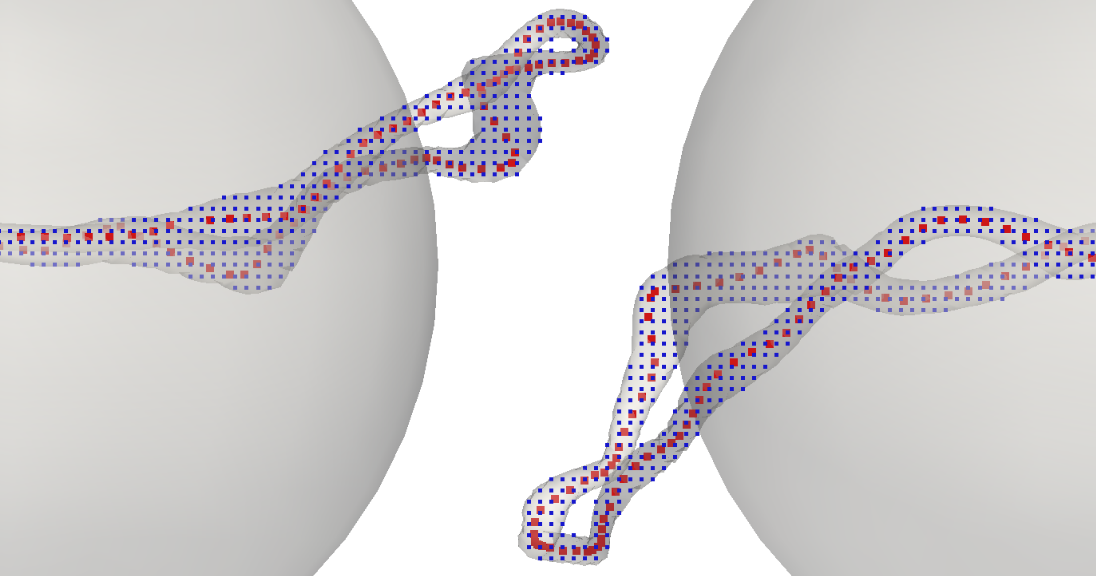}
\end{center}
\caption[Disclination lines around two nanoparticles]
{Disclination lines around two nanoparticles (grey).
Blue small dots correspond to data points with $c_l$ below the threshold
and red large points correspond to points calculated using the `sphere-tracking' (see description in text).}
\label{fig:line_detection}
\end{figure}

The length of a disclination line is of particular interest because it is proportional to its associated energy.
Most of our disclination lines were well-behaved, allowing the use of a ballistic `sphere-tracking' approach which is much faster than connectivity-based approaches.
An arbitrary starting point is chosen, and a second is chosen nearby.
We define the unit vector connecting these two points as our (arbitrary) initial direction of travel, and move a short distance along it.
The average position of the defect line near the destination was calculated by averaging the positions of the 10 nearest points, which was enough to encapsulate the cross-section of the line for our grid size.
The new direction of travel is set to be the vector connecting the previous position and this new average, and this direction is followed from the most recent average to arrive at the new destination.
The process continues until the initial starting point is reached.
The length was estimated by summing the distances between neighbouring averaged positions.
This method proved to be accurate and reliable.
In Fig.~\ref{fig:line_detection} the red points indicated the average positions calculated using this method
for an example snapshot.

\begin{figure}
\begin{center}
\includegraphics[width=0.8\columnwidth]{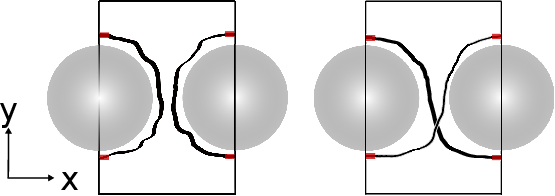}
\end{center}
\caption[Schematic of two nanoparticles and the surrounding disclination lines]
{Schematic of two nanoparticles (grey) and the surrounding disclination lines (black) in the $x$-$y$ plane.
Red regions show the starting areas for the flood-fill to distinguish different defect structures.
Depending on which red regions are connected the defect type was determined.}
\label{fig:sketch_distinguish}
\end{figure}

To accurately distinguish between the different defect types the connections 
between the red regions in Fig.~\ref{fig:sketch_distinguish}
were evaluated using the flood-fill algorithm.
Here the boundaries were treated as fixed.
The connections in conjunction with the number of disclinations and their respective lengths
allow us to determine the defect structure.
For the figure of eight we further distinguished the direction of the twist,
depending on which path crosses over the other.
In over $99.9\%$ of the snapshots this automated analysis was conclusive;
inconclusive exceptions were inspected visually.
%


\section{Results and discussion}\label{s.rd}

\begin{figure*}
\includegraphics[width=0.95\textwidth]{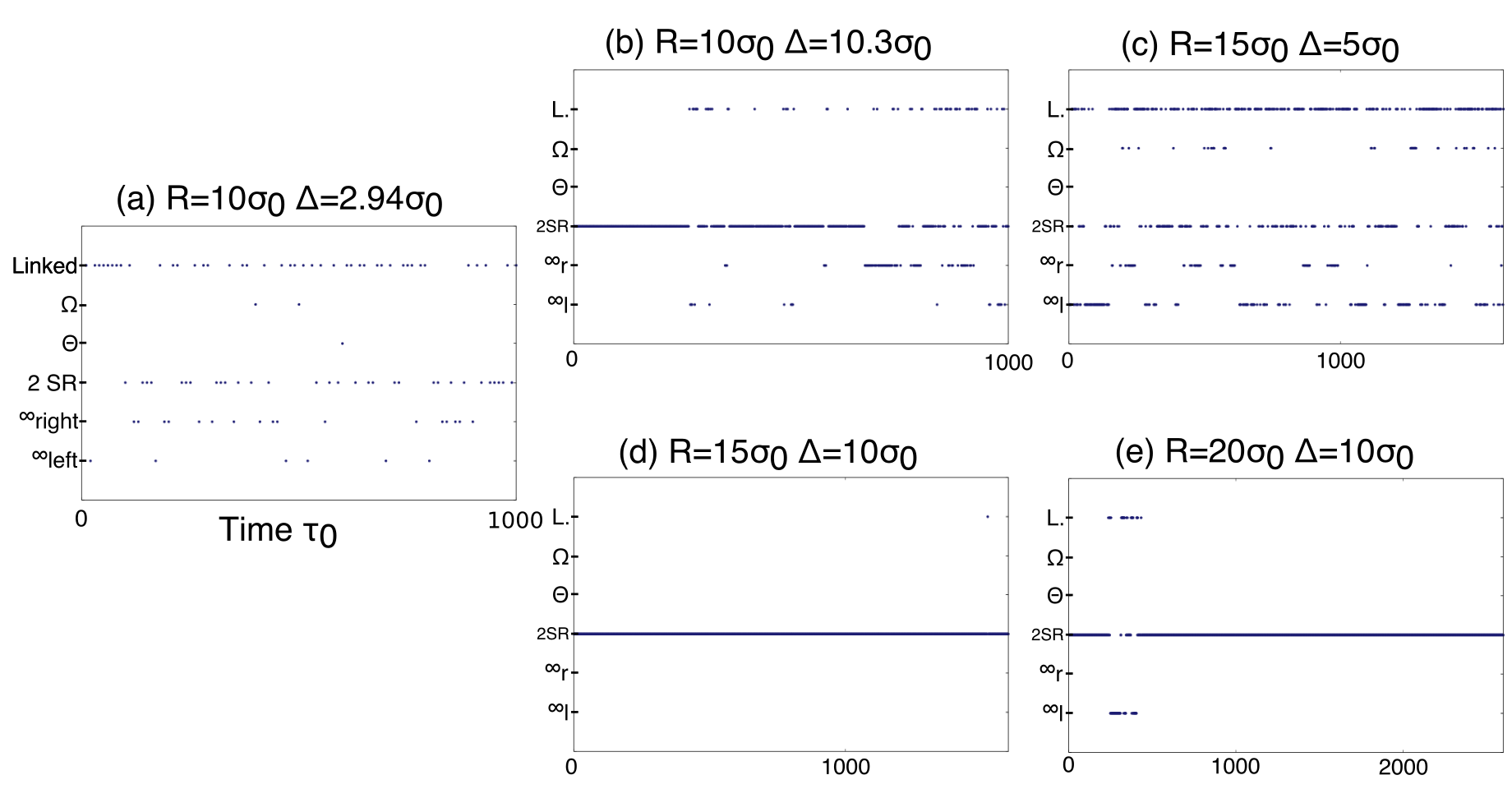}
\caption{\label{fig:results}
Different defect structures shown over time for all five simulations.
Particle radius and surface-to-surface separation are given by $R$ and $\Delta$ respectively.
The defect types are labelled as following: intermediate defect structure (Linked), figure of omega ($\Omega$),
figure of theta ($\Theta$), two separate Saturn rings (2 SR) and figure of eight ($\infty$).
Subscripts indicate right- and left-handed twist.
}
\end{figure*}

\begin{figure}
\centerline{\includegraphics[width=0.99\columnwidth]{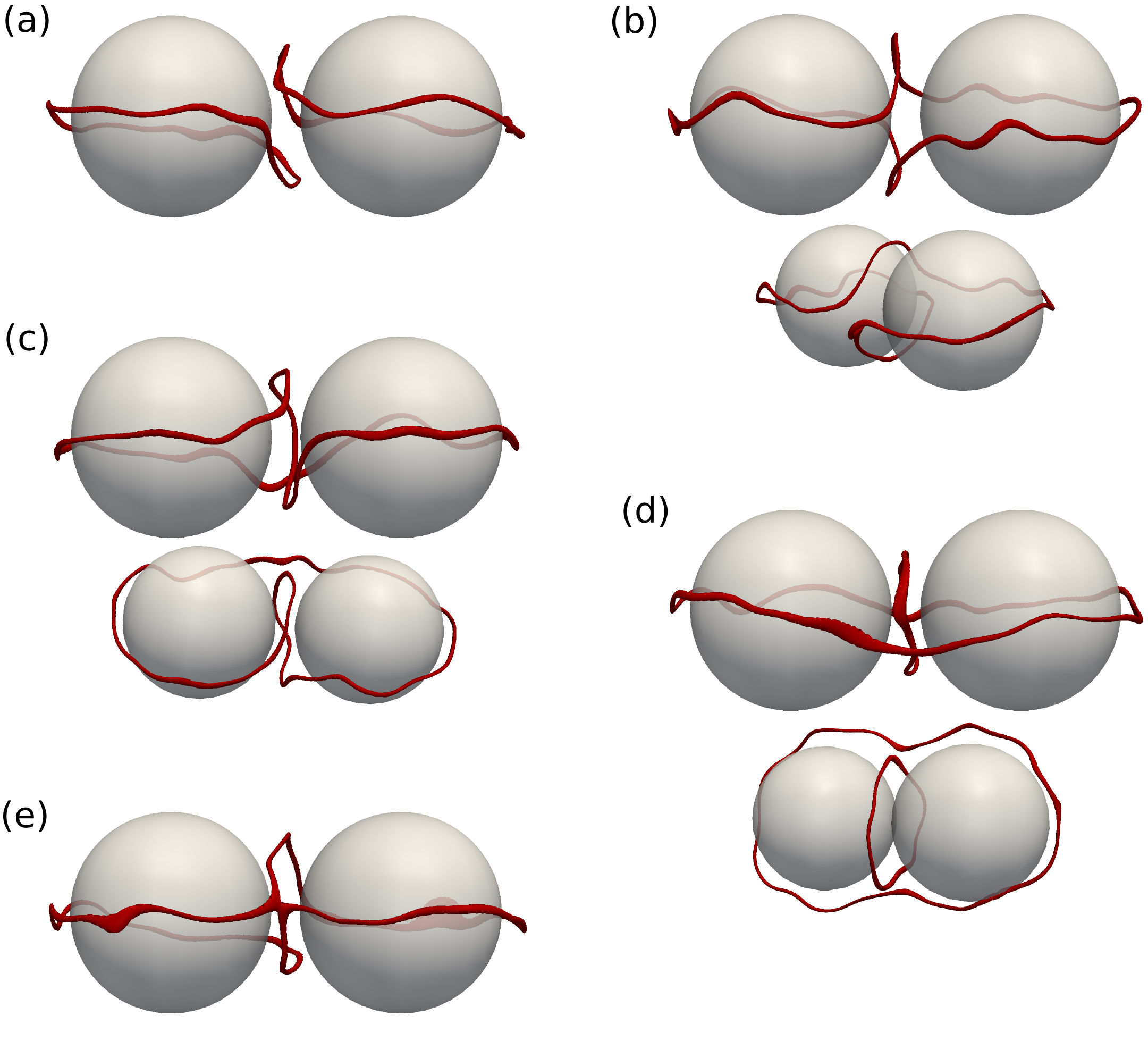}}
\caption
{Defect structures (red) observed for two nanoparticles (grey) of radius $R_\text{NP}=10\sigma_0$ 
in close proximity ($\Delta=2.94\sigma_0$).
Some are shown from two angles for clarity.
(a) two separate Saturn rings (2 SR), (b) figure of eight ($\infty$), (c) figure of omega ($\Omega$), 
(d) figure of theta ($\Theta$) and (e) intermediate defect structure (Linked).}
\label{fig:paraview_entangled}
\end{figure}

In Fig.~\ref{fig:results} the different defect structures over time are shown
for all five systems. 
In total five different structures were observed: 
separated Saturn rings, figure of eight, figure of omega, figure of theta
as well as an intermediate structure, where the two distant parts of the disclination are linked.
With the exception of the intermediate defect structure all these defect structures were also observed in experiments
and using Landau-de Gennes minimisation for micron-sized particles \cite{Ravnik2009e}.
For each of these structures a typical example is shown in Fig.~\ref{fig:paraview_entangled}.

The only non-entangled structure found was two nanoparticles surrounded by a Saturn ring each 
shown in Fig.~\ref{fig:paraview_entangled}(a).
The Saturn rings are strongly bent away from each other to minimise the distortion of the director field, 
which minimises free energy.
The bending effect was also observed for Landau-de Gennes free energy minimisations \cite{Ravnik2009e}.
Fig.~\ref{fig:paraview_entangled}(b--d) show typical snapshots of the figure of eight, the figure of omega 
and the figure of theta from two different angles.
The shape of the defect line is in good agreement with the observations for micron-sized particles in experiments.
Fig.~\ref{fig:paraview_entangled}(e) shows an intermediate structure,
where two distant segments of the disclination line are linked.
For all five structures thermal fluctuations of the disclinations are visible,
since time averaging was avoided.

For nanoparticles with radius $R_\text{NP}=10\sigma_0$ the defect structure transitions frequently between different structures and
no structure persists for longer than $\sim250\tau_0$.
When comparing the results for the two different separations,
one can see that transitions are more frequent for the smaller separation,
suggesting a smaller barrier to interconversion. 
It appears, at least for the time simulated, that two separate Saturn rings are the most frequent structure.
The figure of eight can be seen occasionally, whereas the figure of theta and omega are very rare.
The exact frequencies are given in Table~\ref{table:entangled_occurrences} for all five simulations.
The intermediate linked structure was observed frequently.
This is interesting as this structure is not observed in experiments, suggesting that it is unstable,
which in turn suggests that the particle size does indeed have a significant impact on the energy barriers of the entanglement.

For particles with radius $R_\text{NP}=15\sigma_0$ and a surface-to-surface separation of $\Delta=5\sigma_0$,
entangled defect structures formed, 
and the frequency of transitions is very similar to that for the smaller particles.
With the separation increased to $\Delta=10\sigma_0$ no transitions were observed throughout the production run
and the only structure seen was the two separate Saturn rings.
For particles with radius $R_\text{NP}=20\sigma_0$ and a surface-to-surface separation of $\Delta=10\sigma_0$,
two Saturn rings were observed most frequently. 
For a short period, $\sim 150\tau_0$, the figure of eight with a right-hand twist formed. 

\begin{table}[b]
\caption{\label{table:entangled_occurrences}
Frequency of observations of different entangled defect structures for a different particle radii and separations given in $\%$.
$R_\text{NP}$ and $\Delta$ are the particle radii and their surface-to-surface separation respectively.
Notation for different defect structures as introduced in Fig.~ \ref{fig:paraview_entangled}.
}
\begin{ruledtabular}
\def\arraystretch{1.2}
\begin{tabular}{dddddddddd}
\multicolumn{1}{c}{}&
\multicolumn{6}{c}{Defect structure frequency (in $\%$)}\\
\multicolumn{1}{c}{$\Delta (\sigma_0)$} & 
\multicolumn{1}{c}{2 SR} &
\multicolumn{1}{c}{$\infty_\text{left}$} & 
\multicolumn{1}{c}{$\infty_\text{right}$}  & 
\multicolumn{1}{c}{$\Theta$} & 
\multicolumn{1}{c}{$\Omega$} & 
\multicolumn{1}{c}{Linked} \\
\hline
\multicolumn{7}{c}{$R_\text{NP}=10\sigma_0$}\\
2.94 		& 31.7 & 	5.9 & 	16.8 & 	1.0 & 	2.0 & 	42.6 \\
10.30 	& 71.9 & 	3.6 & 	13.4 & 	0.0 & 	0.0 & 	11.2 \\
\multicolumn{7}{c}{$R_\text{NP}=15\sigma_0$}\\
5.0 & 	24.3 & 	22.2 & 	9.0 & 	0.0 & 	5.2 & 	39.2 \\
10.0 & 	99.88 & 	0.0 & 	0.0 & 	0.0 & 	0.0 & 	0.12 \\
\multicolumn{7}{c}{$R_\text{NP}=20\sigma_0$}\\
10.0 & 	94.5 & 	3.6 & 	0.0 & 	0.0 & 	0.0 & 	1.8 \\
\end{tabular}
\end{ruledtabular}
\end{table}

To summarise, it appears that two separate Saturn rings are the most stable structure
for the nanoparticle sizes studied here.
When the particles are sufficiently close, entangled and intermediate structures were observed
with frequent transitions between them.
The figure of eight was observed to be more stable than the figure of omega
and the figure of theta was the least stable.

At this point it has to be clarified that results presented in Table~\ref{table:entangled_occurrences} 
are indicators only.
Much longer simulation run times would be necessary to give accurate numbers.
For instance, the linked defect structure depends on the resolution grid size chosen when
calculating the weighted order tensor,
i.e.\ a structure may appear to be linked for low resolution, but could be classified otherwise if a higher resolution was used.
We chose a resolution of $0.25\sigma_0$ expecting this to be smaller than the size of the defect core
and hence giving reasonably accurate results.
Furthermore Table~\ref{table:entangled_occurrences} shows that the frequencies
of the figure of eight defect with left-handed and right-handed twist are dissimilar.
This underlines that the simulation run times used were insufficient to obtain accurate results.
Further investigation, for examples of the free energies associated with the different entangled
structures, would have to take this into account;
however it is beyond the scope of this paper.

These concerns aside, our simulations clearly indicate that two separate Saturn rings are more stable
than entangled structures and that the figure of eight is more stable than the figure of omega,
which in turn is more stable than the figure of theta.
This is in quantitative agreement with experimental results for micron-sized particles in strong confinement \cite{Ravnik2007}:
two separate quadrupoles are the only stable structure, observed in $48\%$ of all laser tweezers manipulations.
For entangled structures the figure of eight was found most frequently ($36\%$), followed by the figure of omega ($13\%$).
The figure of theta was observed rarely ($3\%$).
However the binding energies seem to be much smaller for nanoparticles,
indicated by the frequent transitions.
By contrast micron-sized particles' binding energies were calculated to be an order of magnitude 
stronger than the unbound pair and several thousand times stronger than 
particles dispersed in water \cite{Ravnik2007}.

We tried to analyse the relation between the length of the disclination line, the occurrence of a transition
and the state the defect structure is in.
However we found no significant relation.
This is partially due to the high fluctuations in length, where the line stretched or shrank by a few $\sigma_0$
within 500 time steps
and also due to the inaccuracy of the `sphere-tracking' measurement that only provides estimates within $\pm 5\sigma_0$.
Nonetheless it would be an interesting feature to study for more stable entangled defects.

\begin{figure}
\begin{center}
\includegraphics[width=0.48\columnwidth]{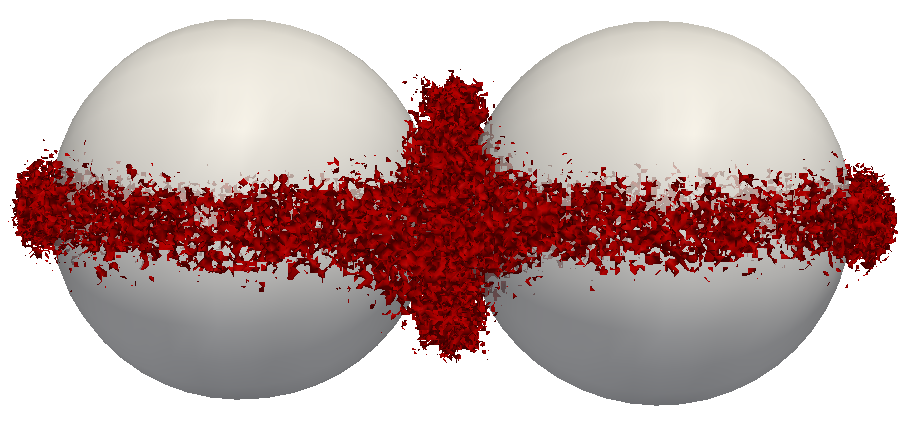}
\includegraphics[width=0.48\columnwidth]{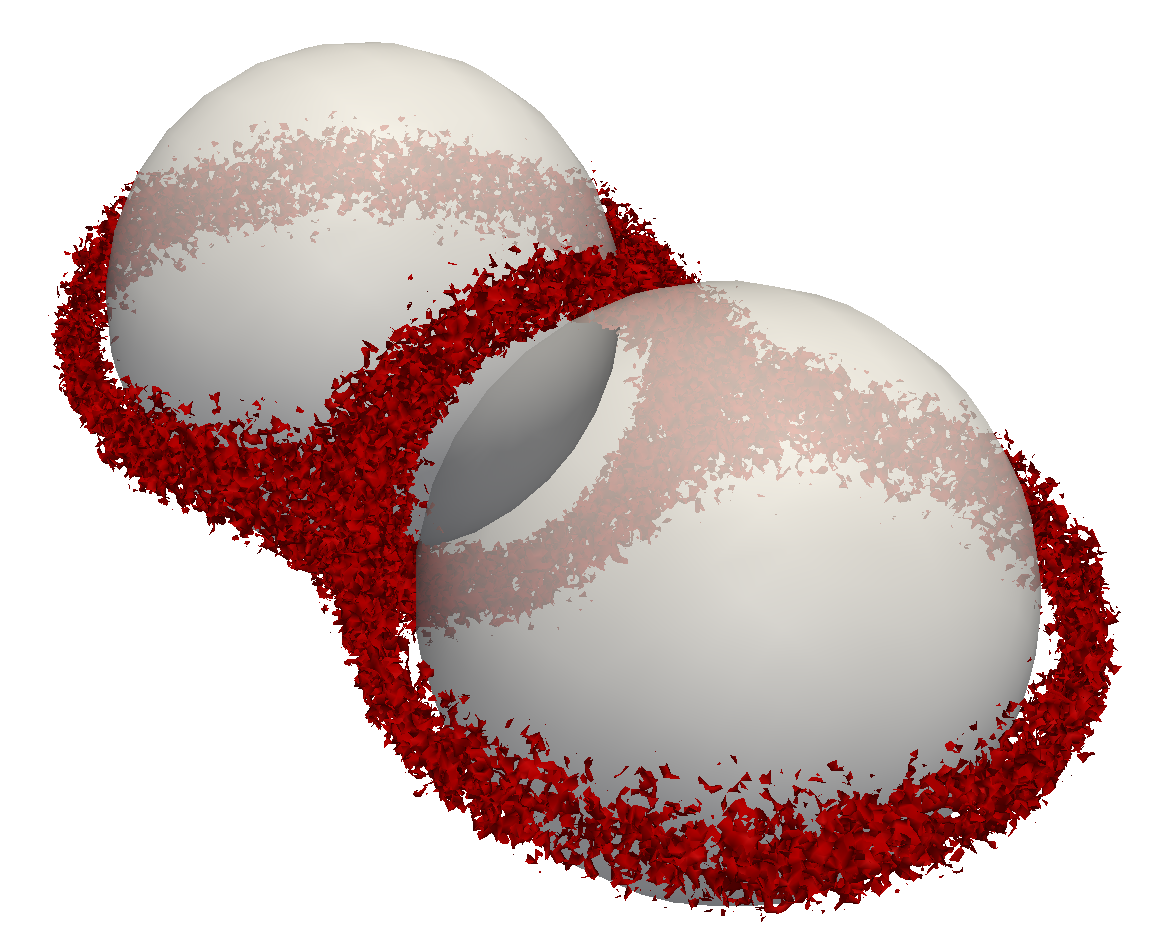}
\end{center}
\caption{Disclination line (red) corresponding to $S<0.4$ time-averaged over entire production run 
for $R_\text{NP}=10\sigma_0$ and $\Delta=2.94\sigma_0$}
\label{fig:three_ring}
\end{figure}

Finally we address the three-ring structure that was observed in previous molecular simulations \cite{Grollau2003,Kim2004,Guzman2003}.
In Fig.~\ref{fig:three_ring} the isosurface corresponding to an order parameter $S<0.4$ is plotted 
for the time-averaged results for the entire production run for $R_\text{NP}=10\sigma_0$ and $\Delta=2.94\sigma_0$.
The time-averaged disclination line forms the three-ring structure with two nodes previously observed.
It appears that it is a product of the frequent transitions and not a stable defect structure itself.
We do not observe such a structure in any instantaneous snapshot.
It appears that the visualisation technique here is important:
the common approach using small bins is inadequate to capture the fast fluctuations.
Using a weighted order tensor including several neighbouring molecules on a fine grid
has proven itself as a very accurate method to locate defect regions.

Our results are in good agreement with calculations by Araki \textit{et al.}\ \cite{Araki2006,Araki2013},
who observed the figure of eight for nanoparticles using numerical methods based on nematodynamics.
In addition we have observed the figure of omega and figure of theta for the first time in molecular simulations.

The results suggest that the entanglement for nanoparticles of sizes studied here is too weak
to observe stable entangled defects, given that the defects frequently transition between different states
and often form intermediate structures which we expect to be highly unstable.
Future work should address larger particle inclusions to generate more stable defect structures;
this could be achieved using the techniques described in Section \ref{s.model}.
Whether a molecular simulation could access the longer length scales 
necessary to observe transitions between stable entangled defects remains to be seen.


\section{Conclusions}\label{s.conclusions}
Molecular simulations were successfully used to simulate defects around two spherical nanoparticles
in close vicinity in a nematic host.
In our exploratory simulations we studied three different radii and several different surface-to-surface separations.
Five different defect structures formed: well-separated Saturn rings, the figure of eight,
the figure of omega, the figure of theta and an intermediate structure.
To our knowledge this is the first observation of the figure of omega and figure of theta for nanoparticles.
The observation of intermediate structures was due to the small size of the particle inclusions
as well as simulations allowing access to much smaller time scales than experiments.
All defect structures observed were qualitatively similar to observations made for micron-sized particles;
however for the particle sizes studied here the transitions were very fast and none of the entangled structures persisted
for more than a few hundred time steps.
This suggests that very small particles cannot be effectively bound together by entangled lines
and that thermal energies are higher than the energy barriers between different entangled defect structures.
Our analysis reveals that the three-ring structure observed in previous molecular simulations 
is solely an artefact of time-averaged superpositions of the different entangled states and not a stable defect structure itself.

To further explore the phenomenon of entangled nanoparticles in molecular simulations 
we suggest a significant increase in particle size.
We expect that the transitions will occur less frequently for larger particles
and hence the dynamics of the system could be studied.
However one has to keep in mind that this would require much longer simulation times
to study transitions.
At this point we have to stress that our simulations were pushing the current limits of computing resources available.
Therefore it is likely that rare event simulation methods 
(see e.g.\ \citep{Bolhuis10,vanErp12})
will have to be applied in larger simulations
to explore all types of entanglement.

\begin{acknowledgments}
The computer facilities provided by Warwick University Centre for Scientific Computing 
and by the ARCHER UK National Supercomputing Service are gratefully acknowledged
as well as the support from the Engineering and Physical Sciences Research Council.
\end{acknowledgments}


\appendix
\section{The Gay-Berne potential}\label{appendix.gb}
The potential originally suggested by Gay and Berne \cite{Gay1981}
is widely used to simulate liquid crystals. 
It can be regarded as a shifted, anisotropic Lennard-Jones potential, 
i.e.\ it depends on the relative orientation of the particles as well as their separation. 
For identical uniaxial particles it can be written as \cite{Berardi1993,Cleaver1996}
\begin{equation}\label{eqn:GB_potential}
U(\uvec{u}_{i},\uvec{u}_{j},\vec{r}_{ij}) 
 =4\epsilon(\uvec{u}_{i},\uvec{u}_{j},\uvec{r}_{ij})
 \bigl[\varrho^{-12} -\varrho^{-6}\bigr]\:,
\end{equation}
where
\begin{equation}\label{eqn:GB_varrho}
\varrho(\uvec{u}_{i},\uvec{u}_{j},\vec{r}_{ij}) = \frac{r_{ij}-\sigma(\uvec{u}_{i},\uvec{u}_{j},\uvec{r}_{ij})+\sigma_{0}}{\sigma_{0}} \:.
\end{equation}
$\uvec{u}_i$ and $\uvec{u}_j$ are unit vectors along the principal axes of the two particles $i$ and $j$,
while $\vec{r}_{ij}=\vec{r}_{i}-\vec{r}_{j}$ is the vector connecting their centres of mass,
$r_{ij}=|\vec{r}_{ij}|$,
and $\uvec{r}_{ij}=\vec{r}_{ij}/r_{ij}$.
$\sigma_{0}$ is a parameter representing the width of the particle. 
$\sigma(\uvec{u}_{i},\uvec{u}_{j},\uvec{r}_{ij})$ is the orientation-dependent range parameter
\begin{align*}
\sigma 
& =\sigma_{0}\bigl[ 1-\tfrac{1}{2}\chi( S_+ + S_-) \bigr]^{-1/2}
\\
\text{where} \quad S_{\pm} &=
\frac{\left(\uvec{r}_{ij}\cdot\uvec{u}_{i}\pm\uvec{r}_{ij}\cdot\uvec{u}_{j}\right)^{2}}{1\pm\chi\uvec{u}_{i}\cdot\uvec{u}_{j}} 
\:.
\end{align*}
Here $\chi$ is given by $\chi= (\kappa^{2}-1)(\kappa^{2}+1)$,
where $\kappa$ is the length-to-width ratio of the particle. 
The strength anisotropy function $\epsilon(\uvec{u}_{i},\uvec{u}_{j},\uvec{r}_{ij})$ 
used in Equation\nobreakspace \eqref{eqn:GB_potential} is given by
\begin{equation}
\epsilon(\uvec{u}_{i},\uvec{u}_{j},\uvec{r}_{ij})=\epsilon_{0}\, \epsilon_{1}^{\nu}(\uvec{u}_{i},\uvec{u}_{j})\, \epsilon_{2}^{\mu}(\uvec{u}_{i},\uvec{u}_{j},\uvec{r}_{ij}) \:.
\label{eqn:epsilon}
\end{equation}
$\epsilon_{0}$ is the well depth parameter 
determining the overall strength of the potential, 
while $\nu$ and $\mu$ are two adjustable exponents
which allow considerable flexibility in defining a family of Gay-Berne potentials. 
$\epsilon_{1}$ and $\epsilon_{2}$ are given by 
\begin{align*}
\epsilon_{1} &=\left[1-\chi^{2}(\uvec{u}_{i}\cdot\uvec{u}_{j})^{2})\right]^{-1/2} \:,
\\
\epsilon_{2} 
&=1-\tfrac{1}{2}\chi' ( S_+' + S_-' ) \\
\text{where}\quad S_{\pm}' &= 
\frac{\left(\uvec{r}_{ij}\cdot\uvec{u}_{i}\pm\uvec{r}_{ij}\cdot\uvec{u}_{j}\right)^{2}}{1\pm\chi'\uvec{u}_{i}\cdot\uvec{u}_{j}} 
\:.
\end{align*}
Here
\begin{equation}
\chi'= \frac{\kappa'^{1/\mu}-1}{\kappa'^{1/\mu}+1} \:,
\end{equation}
where $\kappa'=\epsilon_{\text{S}}/\epsilon_{\text{E}}$ is the ratio of
well depths 
for the side-to-side configuration, $\epsilon_{\text{S}}$,
and the end-to-end configuration, $\epsilon_{\text{E}}$,
of two molecules.

\section{Homeotropic surface anchoring}\label{appendix.homeo}
Instead of using a specific anchoring potential, a simple variation of the standard Lennard-Jones (LJ) 12-6 potential is used.
Here the anchoring is entirely induced by the packing effects of the liquid crystal particles near the surface of the nanoparticle.
For the homeotropic surface anchoring, the Gay-Berne (GB) molecules are allowed to penetrate the surface of the nanoparticle. 
To prevent GB molecules from entirely entering the particle,
a shifted purely repulsive LJ interaction potential $U_\text{homeo}$
\begin{equation}
U_\text{homeo}(\vec{r})
= \left\{
	\begin{array}{ll}
		4\epsilon_{0} \left( \varrho^{-12} - \varrho^{-6} \right)+\epsilon_{0}  & \mbox{if } \varrho^{6} < 2 \\
		0 & \mbox{else}
	\end{array}
\right.
\label{eqn:U_homeo}
\end{equation}
is used.
$\epsilon_{0}$ is an energy parameter chosen to be unity and $\varrho$ is given by
\begin{equation}
\varrho = \frac{|\vec{r}|-\sigma_{c}+\sigma_{0}}{\sigma_{0}} \:.
\label{eqn:varrho}
\end{equation}
Here $\vec{r}_{ij}$ is the vector connecting the positions of the nanoparticle and the GB molecule 
and $\vec{\hat{r}}_{ij}$ is the corresponding unit vector. 
$\sigma_{0}$ is a size parameter and defined as the smallest diameter of the GB molecule; 
in this system $\sigma_0=1$. 
$\sigma_{c}$ is the distance of closest approach between the GB molecule and the nanoparticle, set to
\begin{equation}
	\sigma_{c} = R_{\text{NP}} +\sigma_{0}/2 \:,
\end{equation}
where $R_{\text{NP}}$ is the radius of the spherical nanoparticle.
For this interaction potential, the potential cutoff is chosen to be $R_\text{NP}+1$.

\bibliography{paper}

\end{document}